\documentclass[aps,prb,twocolumn,amsmath,amssymb]{revtex4}

\usepackage{graphicx}
\usepackage{dcolumn}
\usepackage{bm}
\usepackage{color}
\usepackage{cmap}
\usepackage{epstopdf}
\usepackage{xcolor}
\usepackage{physics}

\begin{document}

\preprint{M. O. Ajeesh et al.}

\title{Putative quantum critical point in the itinerant magnet ZrFe$\bm{_4}$Si$\bm{_2}$\\ with a frustrated quasi-one-dimensional structure}

\author{M. O. Ajeesh}
\email{ajeesh@cpfs.mpg.de}
\author{K. Weber}
\author{C. Geibel}
\author{M. Nicklas}
\affiliation{Max Planck Institute for Chemical Physics of Solids, N\"{o}thnitzer Str.\ 40, 01187 Dresden, Germany}

\date{\today}
\begin{abstract}
The Fe sublattice in the compound ZrFe$_4$Si$_2$ features geometrical frustration and quasi-one-dimensionality. We therefore investigated the magnetic behavior in ZrFe$_4$Si$_2$ and its evolution upon substituting Ge for Si and under the application of hydrostatic pressure using structural, magnetic, thermodynamic, and electrical-transport probes. Magnetic measurements reveal that ZrFe$_4$Si$_2$ holds paramagnetic Fe moments with an effective moment $\mu_{\rm eff}= 2.18~\mu_{B}$. At low temperatures the compound shows a weak short-range magnetic order below 6~K. Our studies demonstrate that substituting Ge for Si increases the unit-cell volume and stabilizes the short-range order into a long-range spin-density wave type magnetic order. On the other hand, hydrostatic pressure studies using electrical-resistivity measurements on ZrFe$_4$(Si$_{0.88}$Ge$_{0.12}$)$_2$ indicate a continuous suppression of the magnetic ordering upon increasing pressure. Therefore, our combined chemical substitution and hydrostatic pressure studies suggest the existence of a lattice-volume-controlled quantum critical point in ZrFe$_4$Si$_2$.

\end{abstract}

\maketitle

\section{INTRODUCTION}

Itinerant magnetic systems with low dimensionality and magnetic frustration exhibit enhanced quantum fluctuations leading to the emergence of novel and exotic phases displaying unconventional behaviors. Studying such systems is of great importance as it is becoming increasingly evident that quantum fluctuations play a crucial role in emergent phenomena, including unconventional superconductivity and unconventional metallic phases~\cite{Kotegawa12,Dai15,Inosov16,Chi17,Hirschfeld16,Julian98,Stewart01,Lohneysen07,Shen20}. In order to improve our understanding of these phenomena and the importance of dimensionality and frustration on the ground-state properties investigations of new candidate materials are highly desired.

\begin{figure}[h]
\includegraphics[width=1\linewidth]{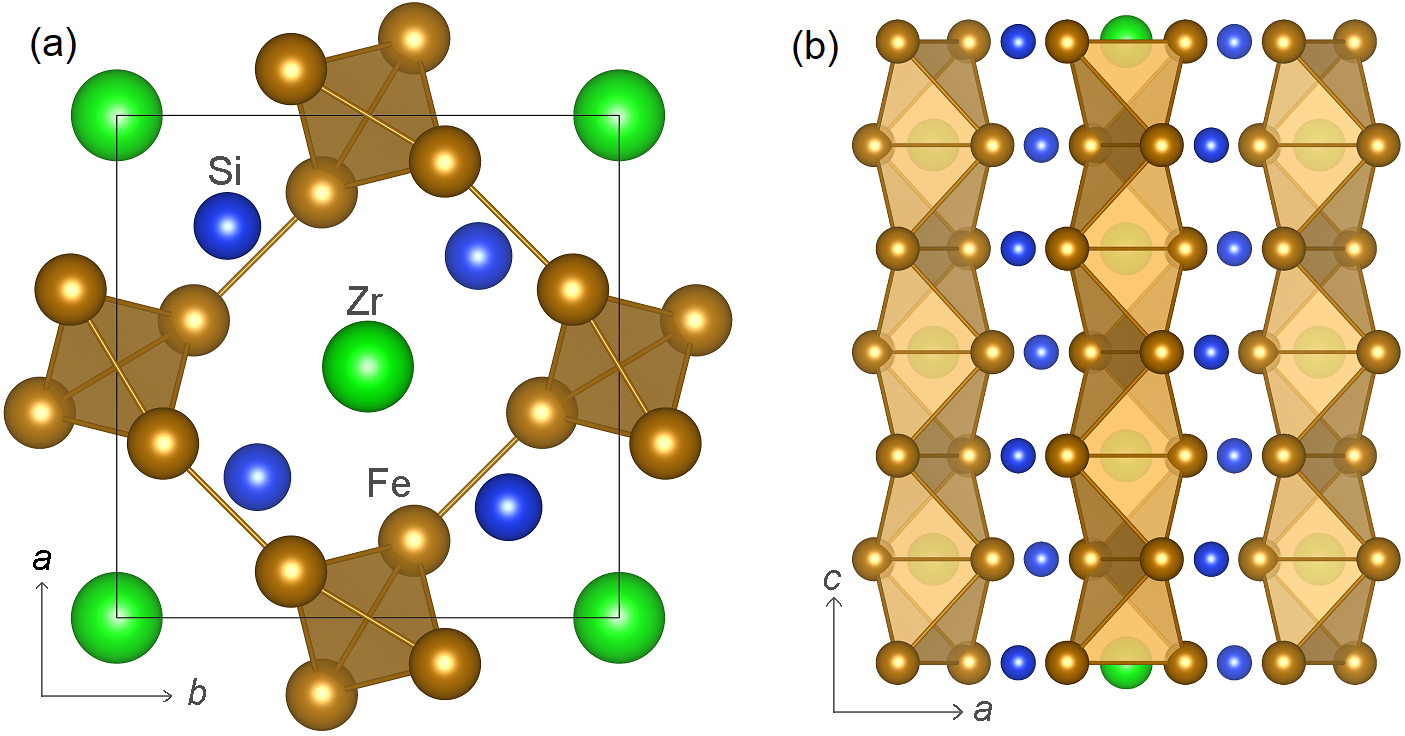}
\caption{(a) Crystal structure of ZrFe$_4$Si$_2$ viewed along the crystallographic $c$ axis. (b) The chainlike arrangement of edge-shared Fe tetrahedra viewed along the $b$ axis.}
\label{CryStr}
\end{figure}

In this regard, ternary intermetallic $A$Fe$_4X_2$ ($A=$ rare earth, $X=$ Si, Ge) compounds are interesting candidates due to their peculiar crystal structure. These compounds crystallize in the ZrFe$_4$Si$_2$-type structure with the $P4_2/mnm$ space group at room temperature~\cite{Yarmoluk75}. The crystal structure consists of slightly distorted Fe tetrahedra, which are edge shared to form chains along the crystallographic $c$ axis, as illustrated in Fig.~\ref{CryStr}. The Fe tetrahedra are prone to magnetic frustration, and the chainlike arrangement provides the quasi-one-dimensional character of the magnetic system, rendering the $A$Fe$_4X_2$ compounds excellent candidate materials to study quantum fluctuations and their effect on the physical properties in low-dimensional frustrated systems.

Previous studies on the $A$Fe$_4X_2$ family mostly focused on compounds with magnetic rare-earth ions $A$. Low-temperature neutron and x-ray diffraction studies on (Er, Dy, Ho, Tm)Fe$_4$Ge$_2$ revealed that the compounds undergo antiferromagnetic ordering at low temperatures resulting in complex spin arrangements due to competing interactions between the magnetic rare-earth and Fe sublattices~\cite{Scho99,Scho00,Scho06a,Scho07,Scho06b,Scho06c,Scho02,Scho04,Scho14,Liu13,Liu15}. In all of the compounds, the magnetic ordering is accompanied by a structural transition from tetragonal to orthorhombic symmetry. $A$Fe$_4X_2$ compounds with nonmagnetic rare-earth elements are even less investigated as only powder neutron diffraction studies on YFe$_4$Ge$_2$, LuFe$_4$Ge$_2$, and YFe$_4$Si$_2$ have been reported~\cite{Scho01,Scho12}. These compounds also order antiferromagnetically at low temperatures with a simultaneous structural transition from tetragonal $P4_2/mnm$ to orthorhombic $Pnnm$ symmetry. While the existing studies address the magnetic structure and the magneto-elastic transitions in these compounds, there are no reports on tuning the magnetic to a non-magnetic ground state by an external control parameter.

In this paper, we present an investigation on a member of the 142 family: ZrFe$_4$Si$_2$. Replacing the rare-earth ions with Zr not only reduces the lattice volume but also changes the valency of the $A$ ion from 3+ to 4+. This may lead to a significant change in the electronic structure and, therefore, of the ground state compared to rare-earth-containing 142 compounds. Here, we studied the ground-state properties of ZrFe$_4$Si$_2$ using magnetic, thermodynamic, and electrical-transport measurements on polycrystalline samples. Our results reveal short-range magnetic ordering below 6~K which is stabilized into a spin-density wave (SDW) long-range order by substituting Ge on Si sites. In addition, we applied external hydrostatic pressure on ZrFe$_4$(Si$_{0.88}$Ge$_{0.12})_2$ to tune the antiferromagnetically ordered ground state toward a nonmagnetic state. Finally, we discuss the temperature--lattice-volume phase diagram in which the magnetic ordering is suppressed by decreasing lattice volume toward a putative quantum critical point (QCP).

\section{METHODS}\label{Methods}

Polycrystalline samples of ZrFe$_4$Si$_2$ were synthesized by a standard arc-melting technique on a copper hearth. At first stoichiometric amounts of the constituent elements (at least 99.9\% purity) were melted in an arc furnace under argon atmosphere, followed by several flipping and remelting of the resulting ingot to ensure homogeneity. Then the as-cast samples were annealed at $1150^\circ$C under a static argon atmosphere for a week. The phase purity of the annealed samples was checked by powder x-ray diffraction (PXRD) using Cu K$_{\alpha}$ radiation and a scanning electron micrograph (SEM). Energy dispersive x-ray (EDX) analysis was used to check the stoichiometry of the samples. SEM studies reveal only a small amount (up to 2\%) of eutectic phase Fe$_3$Si in our samples, enabling us to study the intrinsic properties of ZrFe$_4$Si$_2$. PXRD patterns confirm the tetragonal $P4_2/mnm$ structure type with lattice parameters $a = 6.9916(5)\, {\rm \AA}$ and $c = 3.7551(5)\, {\rm \AA}$, in good agreement with those reported in the literature \cite{Yarmoluk75}. Polycrystalline samples of the substitution series ZrFe$_4$(Si$_{1-x}$Ge$_{x})_2$ were also synthesized following the same procedure. EDX analysis provides the Ge concentrations in the obtained samples as $x=0.12$ (0.1), 0.23 (0.2), 0.34 (0.3), and 0.46 (0.4), where the corresponding nominal Ge concentrations used in the synthesis are given in the parantheses. SEM studies revealed that these samples contain up to 5\% of impurity phases mainly consisting of Fe$_3$(Si$_{1-x}$Ge$_x$).

DC magnetization measurements were carried out in the temperature range between 1.8 and 300~K and in magnetic fields up to 7~T using a superconducting quantum interference device magnetometer (magnetic property measurement sysytem , Quantum Design). The specific heat was recorded by a thermal-relaxation method using a physical property measurement system (PPMS; Quantum Design). The electrical transport experiments were carried out in the temperature range between 2 and  300~K and magnetic field up to 7~T also using a PPMS. The electrical resistivity was measured using a standard four-terminal method, where electrical contacts to the sample were made using $25-{\rm \mu m}$ gold wires and silver paint.

Electrical-resistivity measurements on ZrFe$_4$(Si$_{0.88}$Ge$_{0.12})_2$ under hydrostatic pressure were performed using a double-layered piston-cylinder-type pressure cell with silicon oil as the pressure transmitting medium. The pressure inside the sample space was determined at low temperatures by the shift of the superconducting transition temperature of a piece of Pb. Electrical resistivity was measured using an LR700 resistance bridge (Linear Research) working at a measuring frequency of 16~Hz.

\section{EXPERIMENTAL RESULTS}\label{Results}
\subsection{Physical properties of ZrFe$\bm{_4}$Si$\bm{_2}$}
\begin{figure}[h]
\centering
\includegraphics[width=1\linewidth]{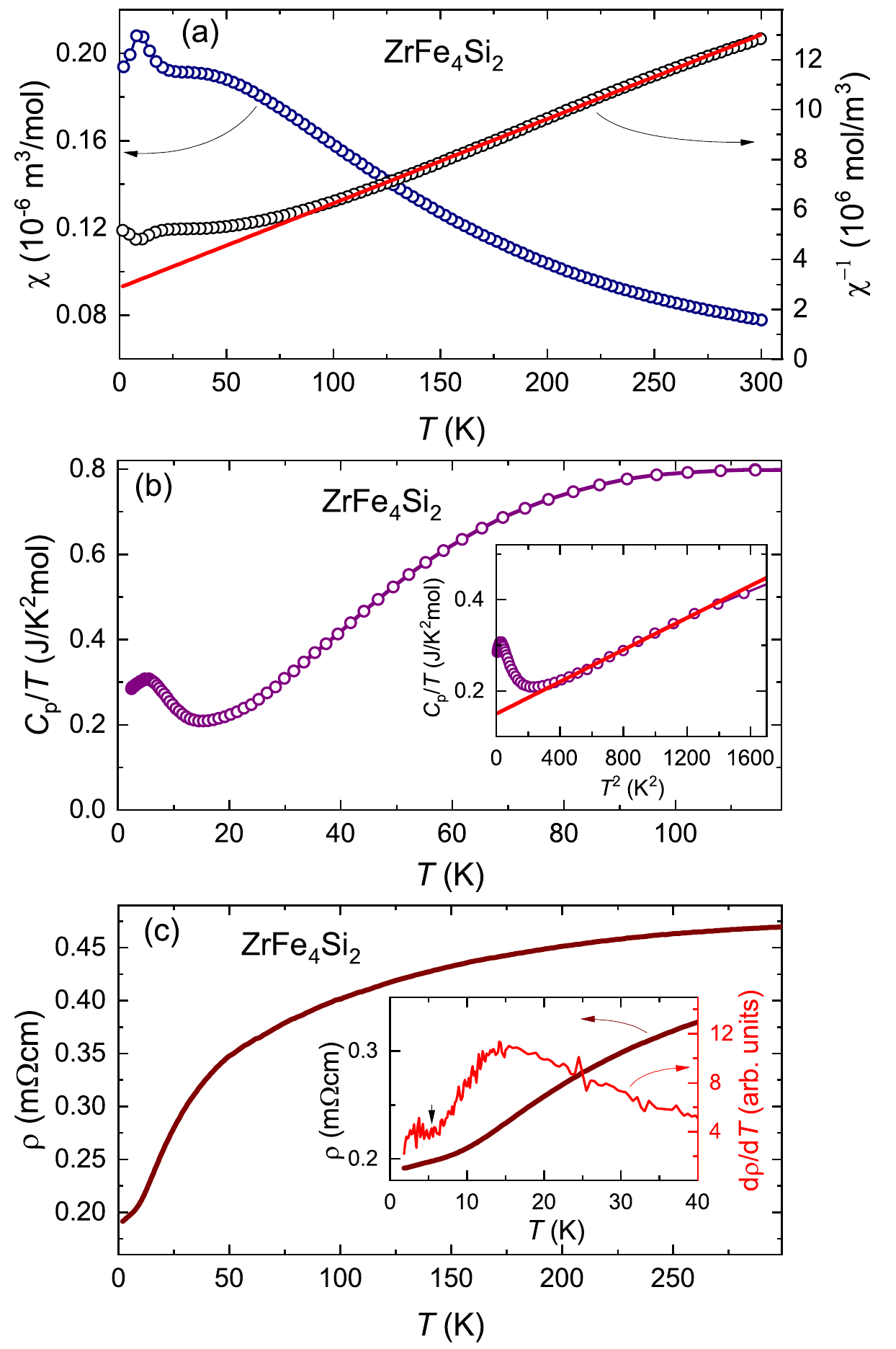}
\caption{(a) Temperature dependence of the DC magnetic susceptibility $\chi(T)$ of ZrFe$_4$Si$_2$ (left axis). The intrinsic susceptibility is obtained by removing the FM impurity contribution, as explained in the main text. The inverse magnetic susceptibility $\chi^{-1}(T)$ is shown on the right axis. The red curve is the Curie-Weiss fit to the data in the temperature interval between 100 and 300~K. (b) Temperature dependence of the specific heat of ZrFe$_4$Si$_2$ plotted as $C_p/T$ vs.\ $T$. The inset shows $C_p/T$ vs.\ $T^2$, where the red line is a linear fit to the data between 20 and 35~K. (c) Temperature dependence of the electrical resistivity $\rho(T)$ of ZrFe$_4$Si$_2$. Inset: an enlarged view of the low temperature region of the $\rho(T)$ curve (left axis) along with the temperature derivative ${\rm d}\rho(T)/{\rm d}T$ (right axis).}
\label{mag}
\end{figure}

In order to understand the ground-state properties of ZrFe$_4$Si$_2$, we have carried out magnetic, thermodynamic, and electrical-resistivity measurements. The temperature dependence of the DC magnetic susceptibility $\chi(T)$ is shown in Fig.~\ref{mag}a. We note that our ZrFe$_4$Si$_2$ samples contain up to 2\% eutectic phase Fe$_3$Si, which orders ferromagnetically above 800~K. Accordingly, this impurity phase induces in the magnetization a ferromagnetic (FM) contribution which, however, saturates at low fields and is therefore field and temperature independent above 1~T and below 300~K, respectively~\cite{Shinjo63}. Thus, the impurity contribution can easily be separated from the intrinsic contribution of ZrFe$_4$Si$_2$. Using magnetization measurements at different fields, this contribution ($M_{\rm FM}$) is found to be rather small with a saturation moment of the order of $1\times10^{-4}{\rm \mu_{B}}/{\rm Fe}$. $M_{\rm FM}$ is then subtracted from the measured magnetization to obtain the intrinsic susceptibility as $\chi(T) = [M(T)-M_{\rm FM}]/H$. At high temperatures, $\chi(T)$ follows a Curie-Weiss behavior $\chi(T)=C/(T-\theta_{\rm W})$, where $C$ and $\theta_{\rm W}$ are the Curie constant and the Weiss temperature, respectively. A Curie-Weiss fit to the $\chi^{-1}(T)$ data (Fig.\ \ref{mag}a, right axis) for $100~{\rm K}<T<300~{\rm K}$ yields an effective moment $\mu_{\rm eff}= 2.18~\mu_{B}$ and a Weiss temperature $\theta_{\rm W}=-85$~K. The relatively large value of the effective moment is a signature of fluctuating Fe moments in the paramagnetic state. Moreover, the negative $\theta_{\rm W}$ indicates that the dominant interactions between the moments are antiferromagnetic. At low temperatures, $\chi (T)$ presents a weak shoulder at around 50~K, followed by a broad peak centered around 6~K. The specific heat data also show a rounded peak at around 6~K, corresponding to the anomaly in $\chi (T)$ (see Fig.~\ref{mag}b). However, there is no evident feature in $C_p(T)$ at $T\approx50$~K, making the presence of any phase transition in this temperature range unlikely. The rounded nature of the anomalies in susceptibility and specific heat at $T\approx6$~K point to short-range magnetic order.

In the inset of Fig.~\ref{mag}b the specific-heat data are plotted as $C_p(T)/T$ vs.\ $T^2$. The linear region observed between 20 and 35~K was fitted with $C_p(T)=\gamma T+\beta T^3$ to obtain the Sommerfeld coefficient $\gamma = 150$~mJ/molK$^2$. This is a very large value for a transition metal compound, indicating very strong electronic correlation effects. Using the $\gamma$ value and the low-temperature susceptibility, we obtain a Sommerfeld-Wilson ratio $R_{\rm W} = (\pi^2k_B^2\chi_{\rm 1.8K})/(\mu_{\rm eff}\gamma)$ of 4.9, which is enhanced compared to $R_{\rm W} = 1$ for the free-electron gas. The enhanced Sommerfeld-Wilson ratio indicates the presence of strong electron-electron magnetic correlations in ZrFe$_4$Si$_2$.

The temperature dependence of the electrical resistivity $\rho(T)$ of ZrFe$_4$Si$_2$ is shown in Fig.~\ref{mag}c. $\rho(T)$ decreases monotonically upon cooling with a strong negative curvature below 100~K, probably originating from strong magnetic correlations. Preliminary M\"{o}ssbauer and muon-spin relaxation ($\mu$SR) experiments indicate the onset of dynamic correlations below 100~K and the onset of weak static magnetic order below~8 K~\cite{Goltz16}. At low temperatures, $\rho(T)$ presents only an extremely weak feature around 6~K, as seen in the temperature derivative of the resistivity ${\rm d}\rho(T)/{\rm d}T$ plotted in the inset of Fig.~\ref{mag}c. Such a weak anomaly in resistivity is also consistent with short-range magnetic order.

It is also important to note that, unlike other compounds in the $A$Fe$_4X_2$ family with trivalent $A$ which show a structural transition associated with the magnetic ordering, temperature-dependent PXRD data do not resolve any structural transition in ZrFe$_4$Si$_2$ around 6~K \cite{Woike11}. Therefore, the low temperature properties of ZrFe$_4$Si$_2$ are not related to a structural phase transition.

\subsection{Tuning the ground state of ZrFe$\bm{_4}$Si$\bm{_2}$ by Ge substitution}

The weak, short-range ordered magnetic ground state in ZrFe$_4$Si$_2$ raises the question of whether the material is situated close to a QCP connected to the disappearance of long-range magnetic order, especially since such long-range order has been observed in other members of the 142 family \cite{Scho01,Scho12}. To investigate this possibility, we have carried out a Ge substitution study. As Ge is larger than Si, varying the Ge content in ZrFe$_4$(Si$_{1-x}$Ge$_{x})_2$ provides a tuning parameter for systematically increasing the unit-cell volume.

To this end, polycrystalline samples of ZrFe$_4$(Si$_{1-x}$Ge$_{x})_2$ with Ge concentrations $x = 0.12$, 0.23, 0.34, and 0.46 were synthesized, and their magnetic properties were studied using various physical probes. We note that our attempts to synthesize samples with $x=0.5$ and 0.6 resulted in phase separation, indicating that compounds with large Ge content are unstable. This is corroborated by the fact that, to our knowledge, pure ZrFe$_4$Ge$_2$ has not been reported in the literature.
\begin{figure}[h]
\centering
\includegraphics[width=0.9\linewidth]{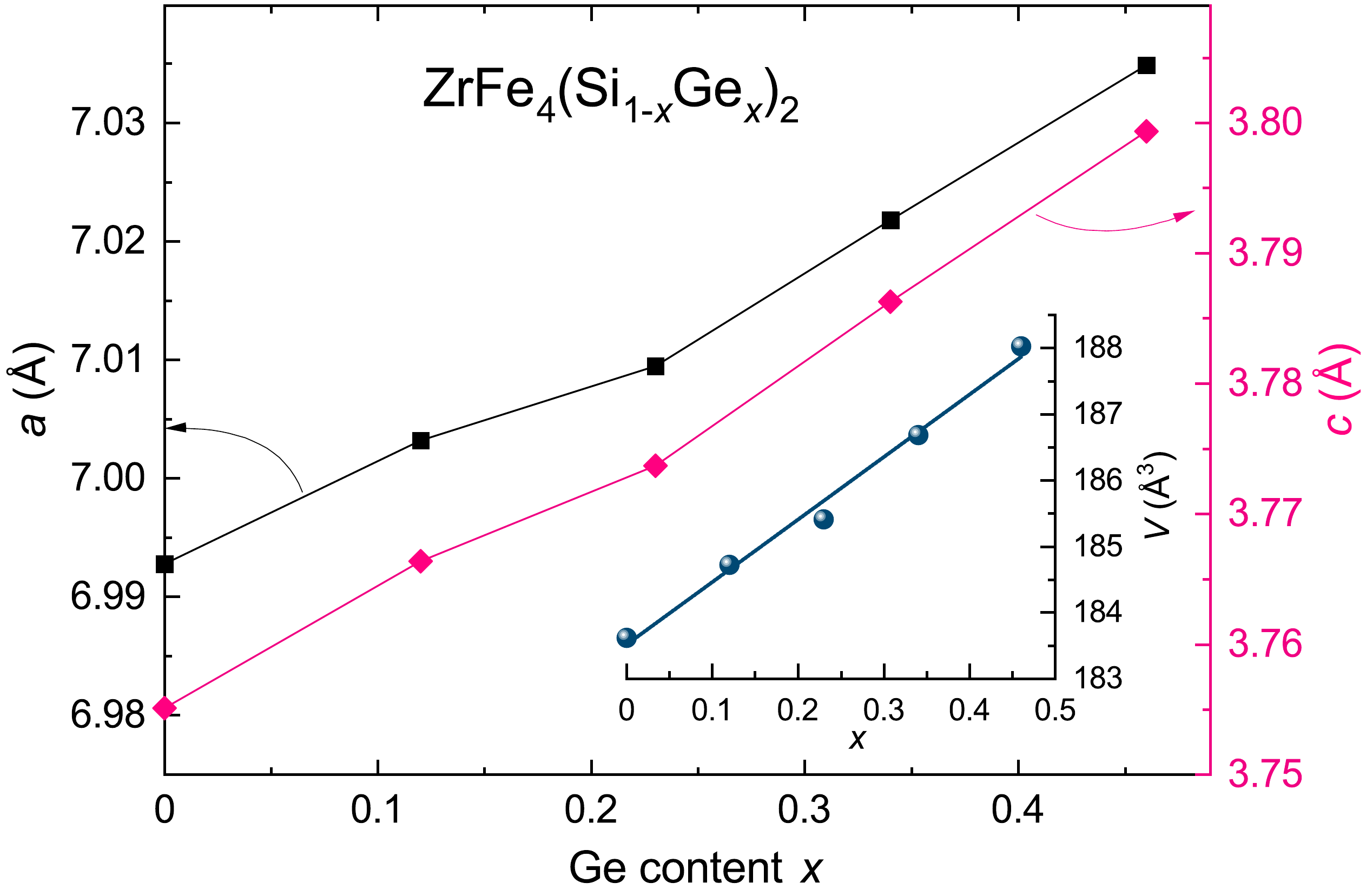}
\caption{Lattice parameters $a$ (left axis) and $c$ (right axis) of the investigated ZrFe$_4$(Si$_{1-x}$Ge$_{x})_2$ samples plotted against their Ge content $x$. The inset shows the change in the lattice volume $V$ with Ge substitution. The solid line in the inset is a linear fit to the data.}
\label{abvsT}
\end{figure}

\begin{figure*}[t!]
\centering
\includegraphics[width=1\linewidth]{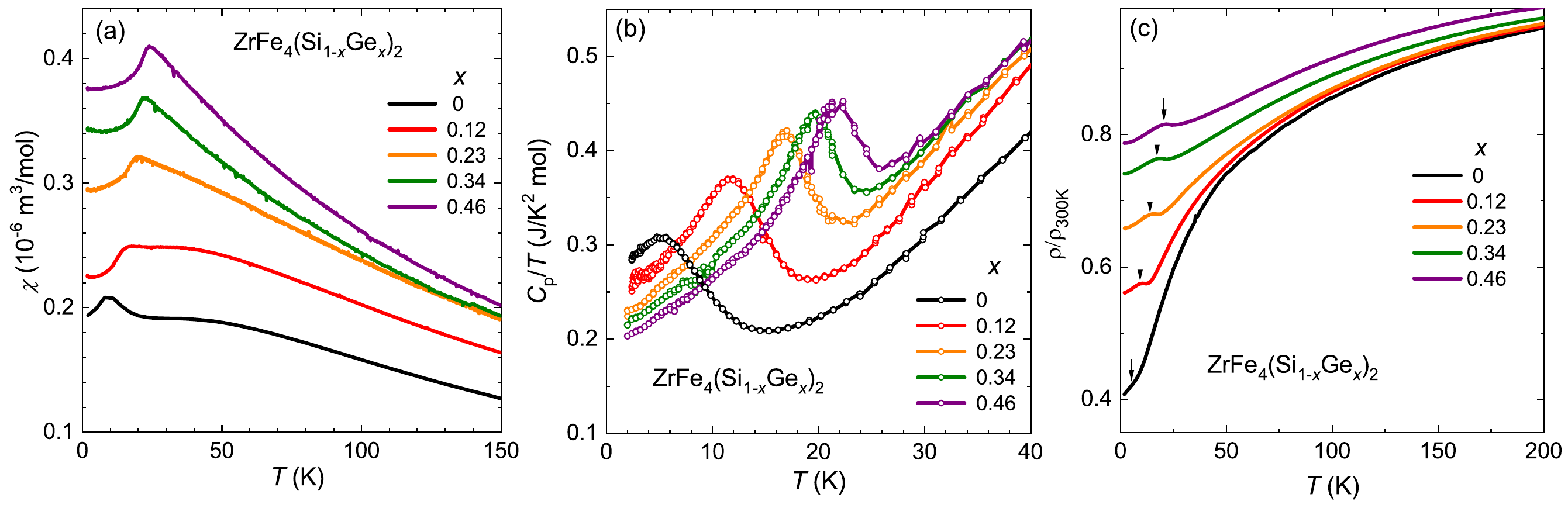}
\caption{Temperature dependence of (a) DC magnetic susceptibility $\chi (T)$, (b) specific heat $C_p(T)/T$, and (c) normalized electrical resistivity $\rho(T)/\rho_{300\rm K}$ of ZrFe$_4$(Si$_{1-x}$Ge$_{x})_2$ for several Ge concentrations. The intrinsic susceptibility
was obtained by removing the FM impurity contribution, as explained in the main text. The position of the peak maxima in the $\chi (T)$ and $C_p(T)/T$ data were taken as the transition temperatures. Corresponding resistive transition temperatures, estimated from the minima in the temperature derivative ${\rm d}\rho(T)/{\rm d}T$, are marked by the arrows.}
\label{Gesub}
\end{figure*}
Figure~\ref{abvsT} depicts the change in the lattice parameters of ZrFe$_4$(Si$_{1-x}$Ge$_{x})_2$ with increasing Ge content $x$, extracted from PXRD measurements. Lattice parameters $a$ (left axis) and $c$ (right axis) monotonically increase with increasing Ge content. Ge substitution with $x=0.46$ results in an increase of $a$ and $c$ by 0.6\% and 1.2\%, respectively. The unit-cell volume $V$ increases nearly linearly, reaching a 2.4\% increase for the compound with $x=0.46$ (see the inset of Fig.~\ref{abvsT}). These results confirm that, as expected, the unit-cell volume of ZrFe$_4$(Si$_{1-x}$Ge$_{x})_2$ continuously increases with Ge substitution.

The physical properties of ZrFe$_4$(Si$_{1-x}$Ge$_{x})_2$ with different Ge concentrations were investigated using magnetization, specific heat, and electrical-resistivity measurements. The temperature dependence of magnetic susceptibility is shown in Fig.~\ref{Gesub}a. The samples in the substitution series contain up to 5\% impurity phases of Fe$_3$(Si$_{1-x}$Ge$_x$), which order ferromagnetically between 800~K and 600~K \cite{Shinjo63,Kan63}. Their temperature independent, saturated magnetization contribution $M_{\rm FM}$ was subtracted to obtain the intrinsic susceptibility $\chi(T) = [M(T)-M_{\rm FM}]/H$. As discussed earlier, $\chi(T)$ of the stoichiometric ZrFe$_4$Si$_2$ sample has a broad shoulder at about 50~K and a small anomaly around 6~K. The shoulder-like feature at 50~K becomes weaker for the $x=0.12$ and 0.23 samples and eventually vanishes for $x=0.34$. The anomaly corresponding to the short-range magnetic order in ZrFe$_4$Si$_2$ shifts to higher temperatures upon increasing Ge content. Moreover, the anomaly develops into a cusp-like feature indicating long-range antiferromagnetic ordering in compounds with larger Ge contents. These results are confirmed by the specific-heat data presented in Fig.~\ref{Gesub}b. The peak in $C_p(T)/T$ shifts to higher temperatures and sharpens with increasing Ge content, with the transition temperature $T_N$ reaching 23~K at $x=0.46$. In the Ge-substituted samples, the anomaly in $C_p(T)/T$ resembles a mean-field-type transition into a long-range ordered phase. We further note that the $C_p(T)/T$ values at the lowest temperatures remain large, in the range of $200-300$~mJ/K$^2$mol. Thus the Sommerfeld coefficient stays large in the whole concentration range, confirming the presence of strong electronic correlations.

The temperature dependent resistivity data $\rho(T)/\rho_{300\rm K}$ provide further details on the nature of the magnetic ordering (see Fig.~\ref{Gesub}c). Already at a low Ge substitution level of $x=0.12$, the $\rho(T)/\rho_{300\rm K}$ curve shows a noticeable upturn around 11~K, reminiscent of a SDW transition. The increase in resistivity is attributed to the formation of an energy gap at part of the Fermi surface due to the SDW formation. With increasing Ge content, the upturn in resistivity becomes much more pronounced and shifts to higher temperatures. These results reveal that Ge substitution stabilizes the weak short-range magnetic order in ZrFe$_4$Si$_2$ into a long-range SDW-type magnetic order.

\subsection{Tuning ZrFe$\bm{_4}$(Si$\bm{_{0.88}}$Ge$\bm{_{0.12})_2}$ by hydrostatic pressure}

The previous results from the Ge substitution in ZrFe$_4$(Si$_{1-x}$Ge$_{x})_2$ study show that application of negative chemical pressure stabilizes the magnetic order in ZrFe$_4$(Si$_{1-x}$Ge$_{x})_2$. This leads to the expectation that external hydrostatic pressure suppresses the magnetic order and eventually drives the system toward an antiferromagnetic QCP. In order to study this, we performed electrical-resistivity measurements under external pressure. We decided to use the slightly Ge substituted compound ZrFe$_4$(Si$_{0.88}$Ge$_{0.12})_2$ for the pressure study since the anomaly in the electrical resistivity of ZrFe$_4$Si$_2$ is only weak and we do not have a well-developed long-range ordered state. In contrast to that, our data for ZrFe$_4$(Si$_{0.88}$Ge$_{0.12})_2$ indicate long-range SDW order and show a clear anomaly in $\rho(T)$ corresponding to the SDW transition, making it an ideal sample for the pressure experiment. The relatively low $T_N=11.4$~K for $x=0.12$ ensures also that moderate pressures will be sufficient to suppress the magnetic order in comparison with compounds with larger Ge concentrations.
\begin{figure}[h]
\centering
\includegraphics[width=1\linewidth]{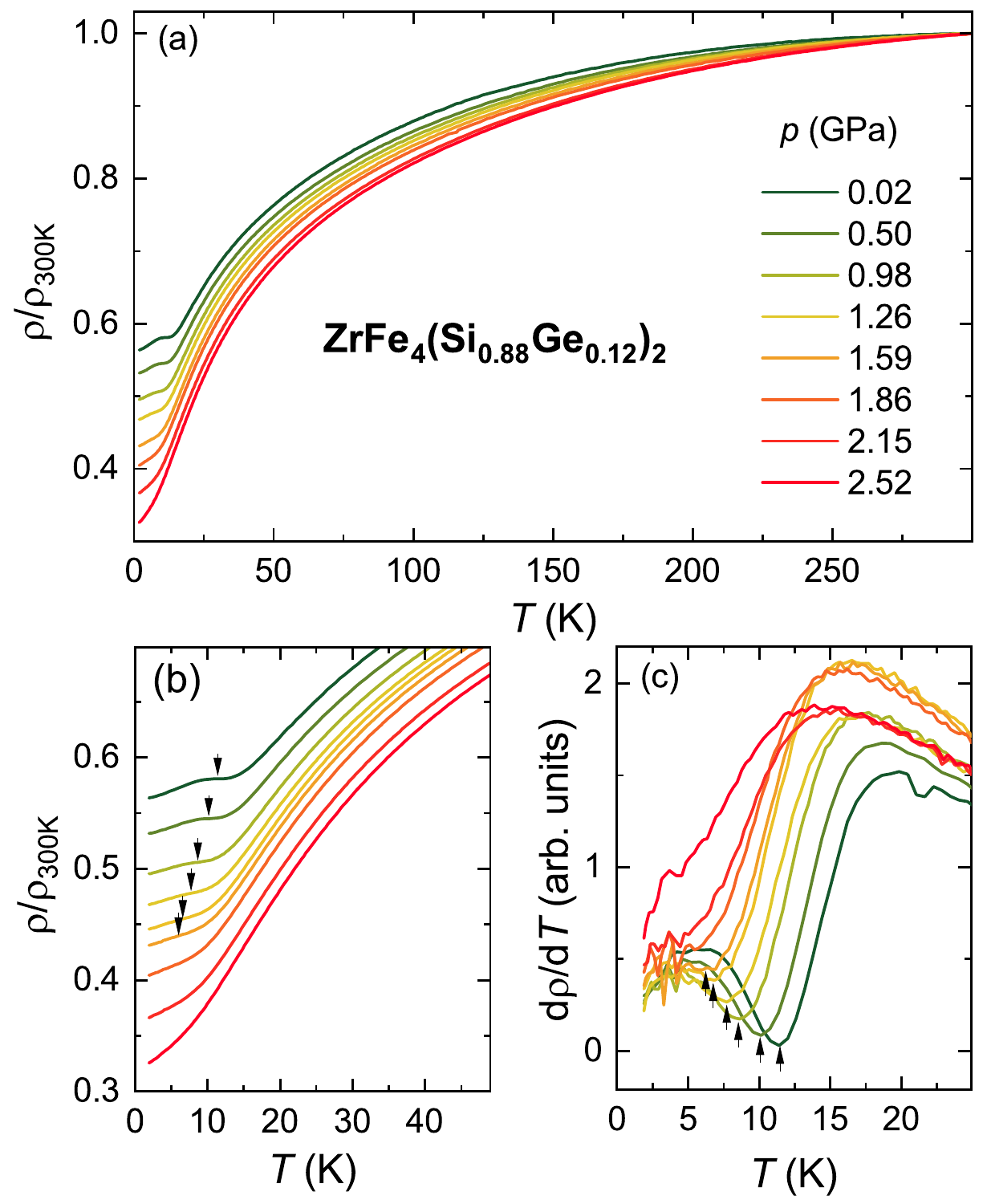}
\caption{(a) Normalized electrical resistivity $\rho(T) /\rho_{\rm 300 K}$ of ZrFe$_4$(Si$_{0.88}$Ge$_{0.12})_2$ as a function of temperature for different applied pressures. (b) Enlarged view of the low temperature region of the $\rho(T)/\rho_{\rm 300 K}$ curves. (c) Temperature derivative ${\rm d}\rho(T)/{\rm d}T$ vs.\ $T$. The arrows indicate the transition temperatures $T_N$ determined by the minima in ${\rm d}\rho(T)/{\rm d}T$.}
\label{rhoP}
\end{figure}

The electrical resistivity of ZrFe$_4$(Si$_{0.88}$Ge$_{0.12})_2$ has been investigated under hydrostatic pressures up to $p=2.5$~GPa and in the temperature range between 2 and 300~K. The curves of the normalized resistivity $\rho(T)/\rho_{\rm 300K}$ for several pressures are plotted in Fig.~\ref{rhoP}a. One immediately recognizes that pressure has a sizable influence on the temperature dependence of the resistivity.  Comparing Fig.~\ref{rhoP}a with Fig.~\ref{Gesub}c, it is obvious that applying pressure has the opposite effect of substituting Ge for Si: The curvature in the temperature range between 20 and 300~K increases with pressure, resulting in a larger slope ${\rm d}\rho(T)/{\rm d}T$ at 20~K (see also~\ref{rhoP}c). Since this strong curvature is very likely connected to the onset of dynamical correlations observed in $\mu$SR~\cite{Goltz16}, applying pressure seemingly strengthens these dynamic correlations. Furthermore, with increasing pressure the anomaly corresponding to the long range order at 11.4~K at ambient pressure shifts to lower temperatures. This can be better seen in Figs.~\ref{rhoP}b and \ref{rhoP}c, where we plot the low-temperature parts of the resistivity and its temperature derivative ${\rm d}\rho(T)/{\rm d}T$. At $p=0.02$~GPa, a small hump in $\rho(T)$ associated with the SDW transition is observed. The transition temperature $T_N$ is determined from the minimum in the temperature derivative of $\rho(T)$. As pressure is increased, the anomaly in resistivity shifts to lower temperatures, as marked by the arrows. Moreover, the magnitude of the upturn strongly reduces with increase in pressure. The anomaly shifts to $5.2$~K at 1.67~GPa. At 1.87~GPa, the anomaly becomes too small and not traceable due to limited resolution of the data in the respective temperature range. At further increased pressures, $\rho(T)$ monotonously decreases upon decreasing temperature without any visible anomaly down to the lowest accessible temperature in our experiments. These results confirm that the SDW transition in ZrFe$_4$(Si$_{0.88}$Ge$_{0.12})_2$ is continuously suppressed to zero temperature by external pressure, which suggests the existence of a pressure-tuned QCP.

\begin{figure}[h]
\centering
\includegraphics[width=1\linewidth]{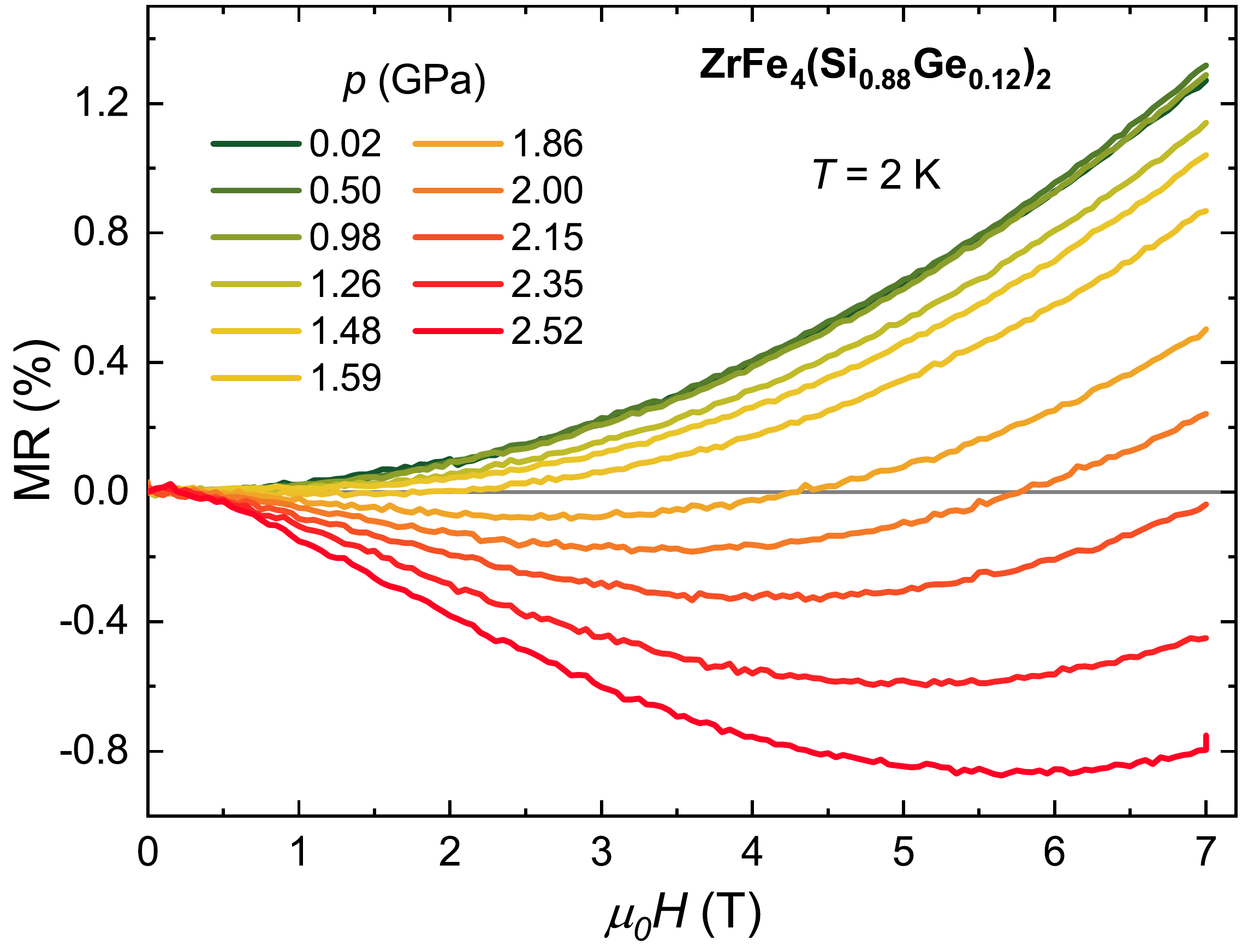}
\caption{Magnetic field dependence of the magnetoresistance ${\rm MR}(H)=[\rho(H)-\rho(0)]/\rho(0)$ of ZrFe$_4$(Si$_{0.88}$Ge$_{0.12})_2$ measured at $T=2$~K under several applied pressures. The gray line corresponds to ${\rm MR}=0$.}
\label{MRvsH}
\end{figure}

The magnetoresistance ${\rm MR}(H)=[\rho(H)-\rho(0)]/\rho(0)$ of ZrFe$_4$(Si$_{0.88}$Ge$_{0.12})_2$ shows marked features connected to the suppression of the magnetic order. Figure~\ref{MRvsH} depicts ${\rm MR}(H)$ measured at $T=2$~K for several pressures. At low pressures, ${\rm MR}(H)$  continuously increases upon increasing field exhibiting a quadratic field dependence, which is typical for a metallic system. For $p\geq1.59$~GPa, ${\rm MR}(H)$ decreases initially upon increasing field, displays a broad minimum and increases again. This contrasting behavior of the MR between the low- and high-pressure regions might be attributed to the enhanced magnetic fluctuations associated with the suppression of the magnetic order. The magnetic field quenches such fluctuations and reduces their scattering contribution, giving rise to the negative MR.

\section{DISCUSSION}

\begin{figure}[h]
\centering
\includegraphics[width=1\linewidth]{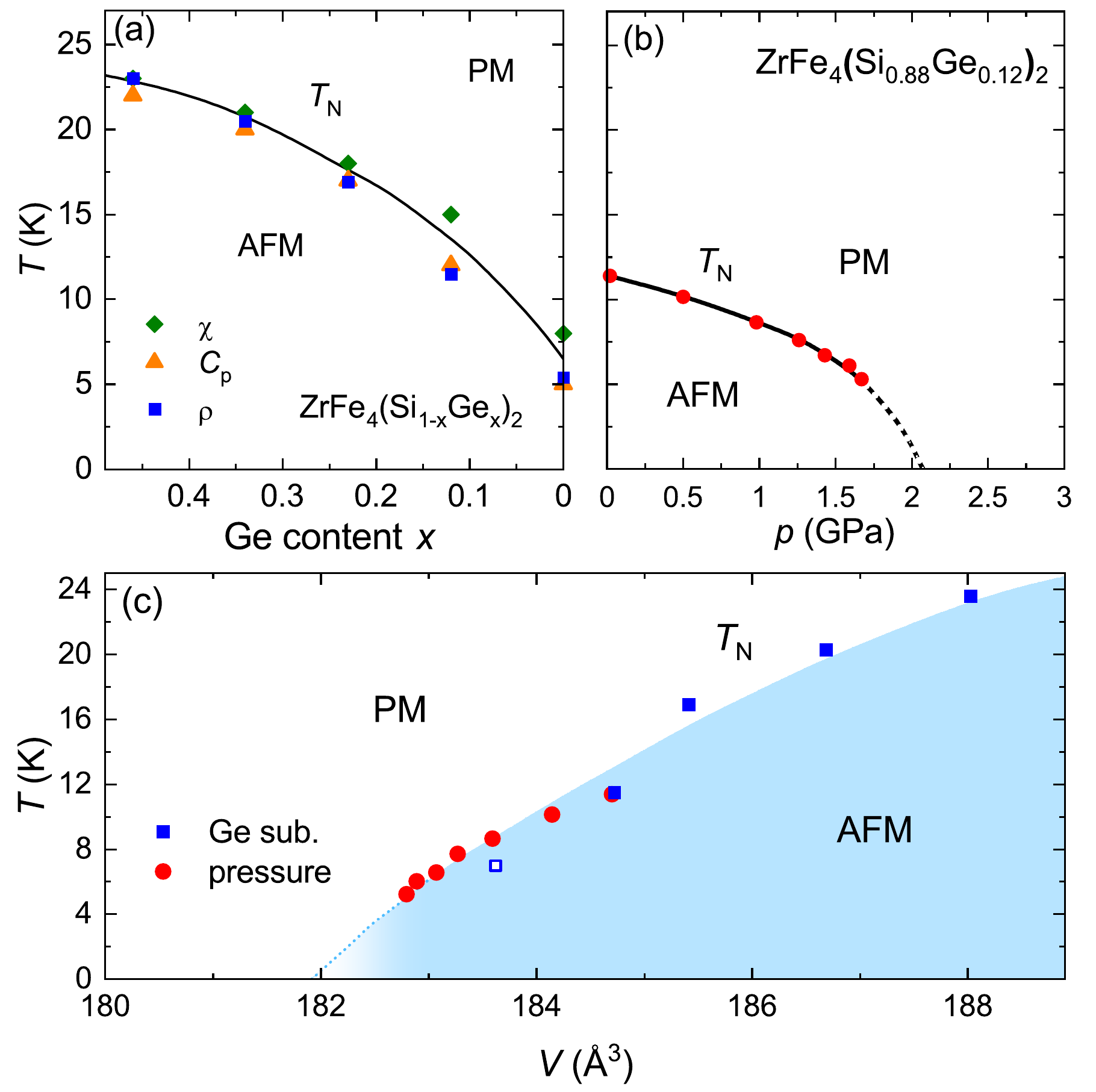}
\caption{(a) Temperature vs.\ Ge-content phase diagram. $T_N$ determined from magnetic-susceptibility, heat-capacity, and electrical-resistivity data are included. (b) Temperature--pressure phase diagram of ZrFe$_4$(Si$_{0.88}$Ge$_{0.12})_2$ determined from the electrical-resistivity data. (c) Temperature--lattice volume phase diagram showing $T_N$ determined from electrical resistivity data from both Ge substitution and high-pressure studies. The open symbol corresponds to the weak anomaly observed in ZrFe$_4$Si$_2$. The solid lines are guide to the eyes. The dashed line is an extrapolation to the experimental data.}
\label{PhaseD}
\end{figure}

The results from the Ge substitution studies in ZrFe$_4$(Si$_{1-x}$Ge$_{x})_2$ are presented as a temperature--Ge-content phase diagram in Fig.~\ref{PhaseD}a. The transition temperatures $T_N$ obtained from magnetic susceptibility, heat capacity, and electrical resistivity data are plotted. Increasing Ge concentration stabilizes the short-range magnetic order present in ZrFe$_4$Si$_2$ into a SDW phase observable in Ge-substituted ZrFe$_4$Si$_2$. $T_N(x)$ is continuously enhanced with increasing Ge content $x$, reaching 23~K at $x=0.46$.

As expected, external hydrostatic pressure produces the opposite effect to that of the negative chemical pressure from Ge substitution. The results obtained from the electrical-resistivity measurements on ZrFe$_4$(Si$_{0.88}$Ge$_{0.12})_2$ are summarized in the temperature--pressure phase diagram presented in Fig.~\ref{PhaseD}b. At ambient pressure, the compound orders antiferromagnetically at $T_N=11.4$~K. With increasing pressure, $T_N$ is monotonously suppressed to lower temperatures, reaching 5.2~K at $p=1.67$~GPa. No traceable anomaly in resistivity can be resolved at higher pressures. However, an extrapolation of the experimental data proposes that the magnetic ordering is suppressed to zero temperature at a critical pressure of $p_c\approx2.1$~GPa, suggesting the presence of a pressure-tuned antiferromagnetic QCP.

In order to compare the effect of Ge substitution and hydrostatic pressure on the magnetic ordering, it is useful to use the unit-cell volume as a common scale. High-pressure PXRD investigations on isostructural LuFe$_4$Ge$_2$ revealed a nearly linear pressure dependence of the lattice volume, yielding a bulk modulus of about 160~GPa \cite{Ajeesh20}. In ZrFe$_4$Si$_2$ a similar pressure dependence of the lattice volume and, therefore, a similar bulk modulus is expected. Thus, we use a bulk modulus of 160~GPa to estimate the lattice volume at different pressures in ZrFe$_4$(Si$_{0.88}$Ge$_{0.12})_2$. The transition temperatures determined from electrical-resistivity data of the Ge substitution series and that of ZrFe$_4$(Si$_{0.88}$Ge$_{0.12})_2$ under hydrostatic pressure are plotted against the unit-cell volume $V$ in the combined temperature--lattice-volume phase diagram shown in Fig.~\ref{PhaseD}c. The change in $T_N$ with Ge substitution and with hydrostatic pressure is rather consistent and evidences the lattice volume as the governing control parameter. Furthermore, $T_N$ shows a continuous suppression of the magnetic ordering with decreasing lattice volume toward a putative antiferromagnetic QCP at $V\approx182$ {\AA}$^3$.

A highly debated question is whether the Fermi liquid behavior expected for a metal at low temperatures breaks down at a QCP, resulting in a non-Fermi liquid (NFL). In the electrical resistivity an NFL is characterized by a deviation from the quadratic dependence of $\rho(T)$ at low temperatures ($\rho=\rho_0+AT^n$, $n<2$), which is expected for a Fermi liquid. Our data in the temperature range down to 2~K seem to indicate a temperature exponent smaller than 2 close to $p_c$ and a recovery of Fermi liquid behavior ($n=2$) at higher pressures. However, these data give only a first hint. The temperature exponent $n={\rm d}\ln(\rho-\rho_0)/{\rm d}(\ln T)$ shows a significant temperature dependence in the low-temperature region for all pressures. In addition, the increased noise in the low-temperature data makes the accurate determination of the exponent difficult. These first results indicate that experiments at lower temperatures are highly desirable.

Magnetic QCPs in transition metal systems are, meanwhile, a well-established research topic. Early cases were the ferromagnetic systems ZrZn$_2$ and NbFe$_2$~\cite{Matthias58,Shiga87}. Presently, the most prominent examples are certainly the Fe pnictides and chalcogenides because there the disappearance of an antiferromagnetic (AFM) state results in the onset of unconventional superconductivity~\cite{Kamihara08,Johnston10}. It is therefore interesting to compare ZrFe$_4$Si$_2$ with well-studied transition metal systems close to a magnetic QCP. The $T$-dependence of the susceptibility, with Curie-Weiss behavior at high temperatures and a leveling out or maximum at lower temperatures is common in transition metal systems close to a QCP. Also the $T$-dependence of the resistivity, with a pronounced negative curvature in the range $20-100$~K is common in such systems. However, there is one property where ZrFe$_4$Si$_2$ stands out in comparison to most itinerant transition metal systems: it presents a huge Sommerfeld coefficient. In transition metal systems, QCPs do not necessarily result in large $\gamma$ values. In the prototypical system BaFe$_2$As$_2$, e.g., $\gamma$ reaches only a value of 5~mJ/molK$^2$ in the stoichiometric system and values of about 25~mJ/molK$^2$ at the substitution-induced QCP~\cite{Hardy10,Bohmer12}. In ZrFe$_4$Si$_2$, the value $\gamma=150$~mJ/molK$^2$ deduced from the high-temperature ($>20$~K) extrapolation (see Fig.~\ref{mag}b) is already one order of magnitude larger, and far above the values typically observed in transition metal systems, even close to a QCP. Furthermore this extrapolated $\gamma$ value obviously misses a large part of the low energy excitations, since the $C_p/T$ value at the lowest investigated temperature of 2~K is significantly larger, about 290~mJ/molK$^2$. The evolution of $C_p/T$ at 2~K as a function of Ge content evidences this value to present a maximum at or near the putative QCP. To our knowledge, within transition metal systems, the value of 290~mJ/molK$^2$ is only surpassed in the compound LiV$_2$O$_4$, which presents a $C_p/T$ value of 420~mJ/molK$^2$ at low temperature~\cite{Kondo97,Urano00,Brando02}. Several mechanisms have been invoked to explain the huge $C_p/T$ value in LiV$_2$O$_4$~\cite{Anisimov99,Pinettes94,Burdin02,Kusunose02,Yamashita03}. All invoke strong geometrical frustration due to V atoms forming a pyrochlore sublattice. Notably, in many of the further itinerant systems presenting a very large $\gamma$ value, there is compelling evidence for strong frustration too, as e.g., in YMn$_2$ ($\gamma = 180$~mJ/molK$^2$)~\cite{Shiga93}, Mn$_3$P ($\gamma = 100$~mJ/molK$^2$)~\cite{Kotegawa20}, and $\beta$-Mn ($\gamma = 70$~mJ/molK$^2$)~\cite{Nakamura97}. There is a second family of transition metal systems showing a large Sommerfeld coefficient, which includes e.g. CsFe$_2$As$_2$ ($\gamma = 184$~mJ/molK$^2$)~\cite{Wu16}, Ca$_{2-x}$Sr$_x$RuO$_4$ ($\gamma = 250$~mJ/molK$^2$)~\cite{Nakatsuji03}, and CaCu$_3$Ir$_4$O$_{12}$ ($\gamma = 175$~mJ/molK$^2$)~\cite{Cheng13}, but there the large $\gamma$ coefficient is suggested to originate from the closeness to a Mott transition. For ZrFe$_4$Si$_2$, the evolution of the resistivity as a function of Ge substitution or pressure makes this scenario rather unlikely since it indicates the system becomes more metallic when approaching the QCP. Thus the huge electronic specific heat observed at low temperature in ZrFe$_4$Si$_2$ compared to values in transition metal systems supports frustration being relevant in ZrFe$_4$Si$_2$. Already in the context of YMn$_2$, Pinettes and Lacroix demonstrated that frustration can strongly enhance the $\gamma$ coefficient close to a QCP in an itinerant system~\cite{Pinettes94}.

\section{SUMMARY}

In conclusion, we have investigated ZrFe$_4$Si$_2$ using magnetization, thermodynamic, and electrical-transport measurements and tuned its ground-state properties by Ge substitution and by application of hydrostatic pressure. In the crystal structure of ZrFe$_4$Si$_2$ the Fe tetrahedra are prone to magnetic frustration, and their chainlike arrangement represent a quasi-one-dimensional magnetic system, a combination which is expected to enhance quantum fluctuations. Despite having large paramagnetic Fe moments ($\mu_{\rm eff}= 2.18~\mu_{B}$) with dominantly antiferromagnetic interactions, ZrFe$_4$Si$_2$ shows short-range magnetic order below 6~K. Ge substitution on the Si sites acts as a negative chemical pressure and stabilizes the short-range magnetic order into a long-range spin-density wave order. By applying hydrostatic pressure on ZrFe$_4$(Si$_{0.88}$Ge$_{0.12}$)$_2$ we continuously suppressed the magnetic order to zero temperature, as shown by the electrical-resistivity data. Therefore, our combined chemical substitution and hydrostatic pressure study suggests the presence of a lattice-volume controlled antiferromagnetic quantum critical point in ZrFe$_4$Si$_2$. In the hydrostatic pressure experiment on ZrFe$_4$(Si$_{0.88}$Ge$_{0.12}$)$_2$ we can infer a critical pressure $p_c\approx 2.1$~GPa and, indeed, magnetoresistance data indicate enhanced magnetic fluctuations associated with the suppression of the magnetic order. Moreover, zero-field resistivity data point to a breakdown of the Fermi liquid description in the vicinity of $p_c$. In comparison to other transition metal systems, ZrFe$_4$Si$_2$ presents a large specific heat at low temperatures, reaching a $C_p/T$ value of 290~mJ/molK$^2$ at 2~K. This large electronic specific heat at low temperatures supports the relevance of magnetic frustration in ZrFe$_4$Si$_2$. Therefore our results evidence ZrFe$_4$Si$_2$ as a strongly correlated electron system with a constellation of interesting properties and thus worth being investigated in depth.

\section*{Acknowledgment}
This work was partly supported by Deutsche Forschungsgemeinschaft (DFG) through the Research Training Group GRK 1621.

\end{document}